S.Sarma, University of California, Irvine.

# SenseDroid: A Context-Aware Information Exchange Framework for Mobile Sensor Networks Using Android Phones

*Abstract*— **Mobile phones and smartphones have evolved to be very powerful devices that have the potential to be utilized in many application areas apart from generic communication. With each passing year, we see increasingly smartphones being manufactured, which have a plethora of powerful embedded sensors like microphone, camera, digital compass, GPS, accelerometer, temperature sensors and many more. Moreover, the ability to easily program today's smartphones, enables us to exploit these sensors, in a wide variety of application such as personal safety, emergency and calamity response, situation awareness, remote activity monitoring, transportation and environment monitoring. In this paper, we survey the existing mobile phone sensing methodologies and application areas. We also formulate the architectural framework of our project, *SenseDroid*, its utility, limitations and possible future applications.**

## I. INTRODUCTION

THERE are a few strong contributing factors that have radically changed the landscape and perspective of mobile phones from being simple computing and communication devices.

The *rapid advancement in technology* has made it possible to manufacture increasingly cheap yet powerful sensors for mobile phones. The embedded sensors that were primarily meant to provide better user experience in mobile phones (ex. ambient light sensors to automatically adjust screen brightness and accelerometer to change display orientation) are now being exploited for more important, life-critical and technically complex scenarios.

Also, the *easiness with which most smartphones can be programmed* is rapidly changing the whole perspective of mobile phone usage. It is no more a very technically challenging task to write an application that uses disruptive sensing to collect, communicate and share a person's real-time activities, keeping track of one's close-proximity environmental conditions and even monitoring one's vitals.

The *scale* at which modern day's mobile applications are distributed by third party vendors, app markets and other sources; and the increasing number of users of such applications is allowing the collection, analysis and dissemination of information across masses and geographies, thereby imparting the ability to exercise massive remote monitoring and control. Mobile sensing is finding applications on all levels of granularity, individual, group and community (including special interest groups) and on a much larger scale involving the entire population of a city, country and across geographies.

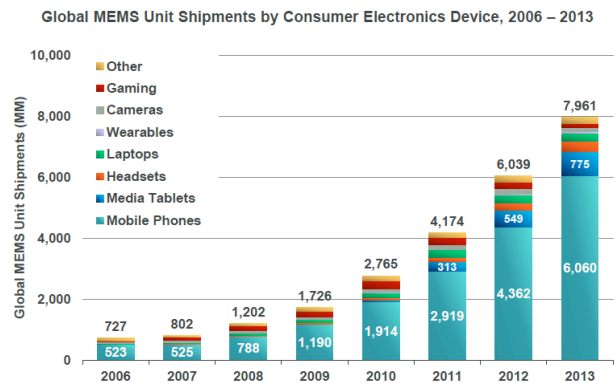

**Figure 1: Mobile phone sensor trends.**

Finally, *the availability of sister-technologies* such as the inexpensive cloud services enables developers to delegate the back end deployment and management to the companies providing such services (such as Amazon, Cloudera etc.) This enables the developers to make use of additional resources and the ability to collect, store and analyze large-scale sensor data with ease.

Hence, these *four* factors vis-a-vis advancement in technology enabling mass manufacturing of cheap yet powerful data, low learning curve and easiness with which these devices can be programmed, proliferation of such devices and applications across massive audience, and the emergence of auxiliary technologies that catalyze smart application development, have revolutionized the research and economics of mobile sensing, broadening the horizon of the application areas where these can be efficiently and effectively used for a variety of purposes.

Although, the culmination of these advances paves our way towards innovative research and development on the mobile

S.Sarma, University of California, Irvine.

sensing front, there exist many *challenges* that need to be addressed. Some widely asked questions are about the limits to which intelligence should be built into smartphones to support sensing without hindering or jeopardizing user's phone experience, how do we come out with a robust infrastructure that can accurately and effectively handle and classify data across a large number of application-users, how do we build applications that exploit the availability of large-scale sensor data shared by applications without any compromise on security and privacy. As can be guessed, this field does include various other active search fields vis-a-vis pervasive computing, data mining, big-data (and fast-data), artificial intelligence and machine learning.

## II. RELATED WORKS

Over the past few years, advances in wireless networking embedded computing and an unprecedented demand for mobility has spurred active research in the field of mobile sensing. Mobile sensing embodies a wide array of correlated technical and semi-technical sub-areas such as *"mobility"* in sensing [4], *sensing scale and paradigms* [3], *context-awareness* using multi-sensor mobile devices [6], *privacy issues* [7], and various *mobile sensor based network architectures*. This section outlines the ongoing research and significant developments in these areas and their contribution towards the new face of mobile sensor driven applications.

Recent work in the NIMS project [2] and underwater sensing [1] has focused on the importance of kinematics and mobility of the sensors (or devices that have sensors mounted on them). When it is not feasible to build a dense sensor network (due to sensor or instrumentation costs or the massive scale of geography to be covered), mobile nodes form the vehicle for sensors, increasing what is called the "span" of the sensor. The mobility can be controlled and the sensor carriers are typically robots that follow a controlled movement pattern to collect data about regions of interest.

The extent to which the user is actively involved in the sensing system [3] dictates two broad paradigms of sensing: *participatory sensing*, where the user actively engages in the sensor activation and data collection activity and *opportunistic sensing* [5], where the data collection is fully automated with negligible (to nil) user involvement. These two paradigms have their own pros and cons. Opportunistic sensing lowers the burden placed on the user, and since human involvement is not a necessity, the only requirement for the application to fulfill requirements is to be active on the user's device. Furthermore, this kind of sensing is inevitably required for some applications that demand massive data accrued over a period of time (ex. community sensing). But this approach does not use "the" most important resource, human participation. One of the main challenges of using opportunistic sensing is the phone context problem. There is nothing much an application can do by itself; if for example, the application wants to take a sound sample from a high-traffic subway "only" when the phone is out of the pocket or bag.

The use of context in mobile devices is receiving considerable attention in various fields of research including mobile computing, wearable computing, augmented reality, ubiquitous computing and human-computer interaction [8]. The motive is to augment mobile devices with awareness of their environment and situation as context. The actual utility of context-awareness in mobile systems has been demonstrated in a wide range of application examples, in obvious domains such as fieldwork and tourism as well as in emerging areas like affective computing based on bio-sensing [9]. Also, it has been shown, that context is useful at different levels within a mobile device. At systems level, it can be exploited for example for context-sensitive resource and power management. At application level, context-awareness enables both adaptive applications and explicitly context-based services. And at the user interface level, the use of context facilitates a shift from explicit to implicit human-computer interaction, toward less visible if not invisible user interfaces [9, 10]. There have been rapid advancements in the new sensing paradigm of opportunistic people-centric sensing (leveraging humans as part of the sensing infrastructure) which is giving rise to new problems especially in the field of security and privacy [7].

Several projects taken up by university students as wells as by industry researchers are related to our work of exploring the sensors in mobile phones. *SensorPlanet* is a Nokia-initiated global research framework for mobile-device centric wireless sensor networks [11, 12]. It provides hardware platforms and a research environment that helps researchers collect sensor data on a large and heterogeneous scale and establishes a central repository for sharing the data. *SenseWeb*[13] a Microsoft Research sponsored project, provides shared sensing resources and sensor querying and data-collection mechanisms to develop sensing applications.

The UCLA Urban Sensing [14] initiative has a vision of equipping users to compose a sensor-based recording of their experiences and environment by leveraging sensors embedded in mobile devices and integrating existing public outlets of urban information (such as weather, traffic, and air quality). Urban Sensing is exploring how to coordinate these individual stories of everyday life to document the urban environment, as well as how to fuse them with other sensed data about the city and feed that back into the physical, collective experience in urban public spaces. The faculty and students of Dartmouth College worked on a project called *MetroSense* [15], which explored the area of people centric sensing using the mobile phones in an urban environment.

To date, very few published papers present a network architecture solution for general-purpose mobile sensors network. Proposals in the areas of tiered sensor networks, delay tolerant sensor networks, and sensor network and ubiquitous computing middleware architectures represent the most closely related work. A number of projects propose the use of delay-tolerant networking within the context of sensor networks [16, 17]. An active research is being carried out in exploring architectures that deal with fault tolerance for mobile sensor networks for pervasive information gathering [17]. The construction and deployment of sensor networks presents new types of challenges. Generally a DTN-based approach is taken to deal with them [16].

S.Sarma, University of California, Irvine.

| Project Title / System | Hardware Description | Software Description | Communication Modules | Security Method/Techniques | Applications | Area |
|---|---|---|---|---|---|---|
| AnonySense | Nokia N800, Apple iPhone | SQL, XQuery, Linux server, C++, XML, Open SSL | Bluetooth, WiFi | HAs-1, RSA, Open SSL secure channel | RogueFinder, ObjectFinder | Special Purpose Application |
| Secure Social Aware | Macbook Pro notebook | Mac OS X 10.5 J8E, J2SE 5.0, MIDP 2.0, BlueCove J5Rr-82 Emulator tool, Open Source SimpleJPA tool, VLC media player | Bluetooth | HTTPS, SHA-1 cryptographic hash function | SocialAwareFlicks | Social Interaction |
| SmokeScreen | Nokia 6670, Sharp Zaurus SL-5600 PDA, Nokia 6600 | Symbian OS, Linux 2.4.18, Python, C++, Postgre SQL 7.4.7, Pretec CompactFlash | Bluetooth, WiFi | RSA, AES | Presence-sharing Application | Social Interaction |
| Serendipity | Nokia Series 60 phone | Mobile Information Device Profile MIDP 2.0, MySQL | Bluetooth, WiFi, GPRS | Privacy Control Policies for Users Proximity | BlueAware, BlueDar | Social Interaction |

Much architecture does not handle mobility and thus are not adequate to handle mobile sensing [18, 19, and 20]. These tiered architectures face issues when dealing with large-scale static sensor networks. Tenet [19] is an architectural approach, which has strict tiering. Such strict tiering makes the architecture rigid and is inefficient. Also, they are all targeted at static sensor networks, and are thus inadequate to handle the dynamics of mobility in the sensing environment.

## III. MOBILE SENSORS

With every passing generation (typically six months), mobile phones add richer functionality, which is often paired with introduction of new sensors. For example, accelerometers were initially introduced to enhance user interface. Over a period of time, the use of accelerometers was expanded to enhance the usage of camera, and to automatically determine the orientation of the phone, and rotate the display accordingly, making use of this sensor-based information.

Typically today's phone sensors include a wide variety of sensors, like a *gyroscope, compass, accelerometer, proximity sensor, ambient light sensor, dual cameras (front and back),microphone, GPS, WiFi, and bluetooth radios*. Some sensors are used to *augment* the functionality of already existing sensors, such as using a compass to bolster the GPS functionality. Context recognition is achieved using the light and proximity sensors. The auto-adjustment of display brightness makes use of the *light sensor*, which helps save a lot of power. Proximity sensors are used for a variety of purposes, for example identifying the closeness of phone to the user's, and thereby using this information to infer that the user is trying to speak, and then deactivating the touchscreen and keypad to prevent accidental key press.

The *GPS sensors* allow the phone to be local-aware, enabling a variety of applications ranging from navigation to local search. Very often, the presence of other sensors such as the compass and gyroscope augment the GPS senor by providing increased awareness, with regards to its direction and orientation, enhancing location-based applications. Furthermore, distinct patterns of data collected through these sensors can be used to recognize different activities (ex. an accelerometer data can be used to determine different activities such as walking, running, sleeping etc.) The combination of accelerometer data and a stream of location estimates from the GPS can recognize the mode of transportation of a user, such as using a bike or car or taking a bus.

*Cameras* and *microphones* are powerful sensors, perhaps the most commonly found as well. The information gathered from these sensors can be used in conjunction to determine a lot of things about the users of the phone. For ex. Sound patterns can be classified and used to determine if the user is having a conversation or listening to music. Also, exploiting the front camera to track user's eye movement has led to significant research in developing user interfaces that do not need to be touched at all, and can be activated using the sensor-data alone.

Our project *SenseDroid*, is aimed at demonstrating the usage of some of these sensors to gather data about people, their contexts and environments.

## IV. APPLICATION AND USE CASES

Utilizing mobile sensors to collect analyze and infer data have numerous existing and future applications. Collecting low level sensor data and using it in conjunction with high level events, context, and activities and deriving solid inferences from them, is a field that is being actively researched on. They are being explored in academic as well as industrial circles alike. This section deals with some existing and possible use cases and applications of mobile sensing, data collection and aggregation and inferences drawn from them (decision building).

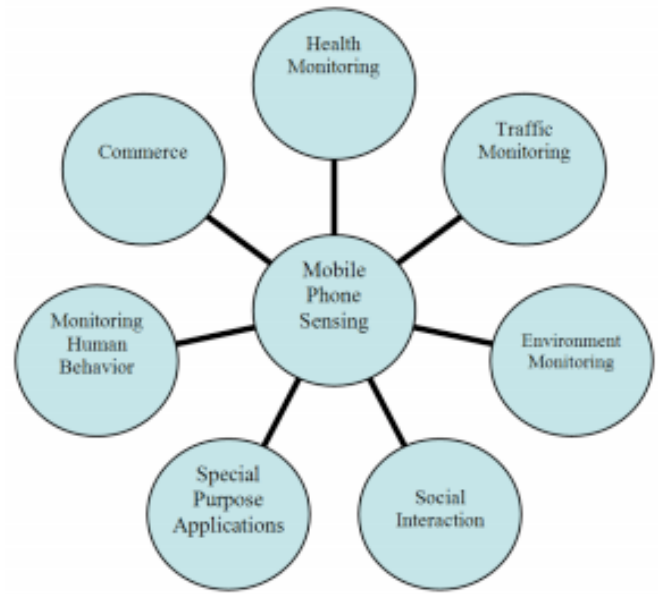

**Figure 2: Application of mobile phone sensing.**

### A. Social Behavior

Having millions of mobiles running a sensor-based application (like *SenseDroid*) can prove to be of immense importance to determine social behavior. This application would gather GPS estimates from millions of users (running this application) over a period of time and store it in a remote

S.Sarma, University of California, Irvine.

**Table 1: Survey of the Sensors in Mobile Phones**

| Model | GPS | Camera | Accelerometer | Gyro | Compass | Proximity sensor | Microphone | Ambient light sensor | Power / Battery |
|-------|-----|--------|---------------|------|---------|------------------|------------|----------------------|-----------------|
| Samsung Galaxy S | Yes with A-GPS support | 5 MP | Yes | No | Yes | Yes | Yes | Yes | Standard battery, Li-Ion 1500 mAh |
| Samsung Galaxy S II | Yes with A-GPS support | Back 8 MP camera, front 2 MP camera | Yes | Yes | Yes | Yes | Yes | No | Standard battery, Li-Ion 1650 mAh |
| Motorola Droid A855 | Yes with A-GPS support | 5 MP | Yes | No | Yes | Yes | Yes | Yes | 1400mAh Internal rechargeable removabe lithium-ion battery |
| Samsung Google Nexus S | Yes with A-GPS support | 5 MP | Yes | Yes | Yes | Yes | Yes | No | Standard battery, Li-Ion 1500 mAh |
| HTC One X | Yes with A-GPS support | 8 MP | Yes | Yes | Yes | Yes | Yes | No | Standard battery, Li-Po 1800 mAh |

server. The data hence gathered can be used to determine or predict a lot of important events. For ex. Assume that an application uses GPS data from millions of users in a city on a regular basis and stores this information in a central server. This data can be used to determine the sections of population that go to a particular event on a regular basis (ex. some jazz event or a football match). The sensors in an individual's cell phone can be used to collect data and store it on a server; such data can be used to generate patterns, which in turn can be used to predict an occurrence of an event in the future, based on historical data and pattern. Such inferences can be further used to predict traffic flows on special events and hence helping the concerned personnel to be prepared in advance.

*B. Transportation*

Mobile phone sensing systems can be used to provide fine-grained traffic information on a large scale using mobile phones that facilitate services such as accurate travel time estimation for improving commute planning. Such information can be used by the transportation departments for numerous uses, ranging from traffic re-routing to identifying roadblocks. Several projects such as the *MIT VTrack* and *Mobile Millennium project* are working towards this direction actively.

*C. Environmental Monitoring*

Conventional methods of measuring environmental pollution rely on collecting data and aggregating statistics pertaining to a community, over a period of time. Active research is going on to use sensors in phone that enable sending and receiving messages to and from users, on a usual basis to gather the data about an individual's footprint (carbon and environmental footprint) and notify users of their impact on the environment and/or their exposure to the amount of pollution that they have been exposed to. *UCLA's PEIR project* is one such research that uses sensors in phones to inform users of their impact and exposure to environment.

*D. Personal and Community Health Monitoring*

Most of the information about an individual's or community's health comes from data collected from hospitals, infrequent doctor consultation and self-report surveys. Sensor-enabled mobile phones have the potential to collect continuous sensor data that can dramatically change the way health and wellness are assessed as well as how care and treatment are delivered. For example, it is not very difficult to collect information about a user's activities and relate this information to personal health goals and notify or give a simple feedback to the user on a regular basis. Hence this application can be used to prevent poor behavior patterns and encourage physical activity. *The UbiFit Garden*, a joint project between Intel and University of Washington is researching on using mobile sensors to generate fitness awareness.

*E. Location and Context-Aware Emergency Response*

Ability to activate sensors remotely and thereby getting information on-the-fly can be effectively used for emergency situation reaction. For example, the temperature sensors in mobile phones can be used to determine the temperature and whenever a predetermined threshold is reached, the data can be used to send an alert to concerned personnel (or a fire brigade). Similarly the sensor data from a microphone, camera and accelerometer can be used in conjunction to identify if a fast moving user has come to a sudden halt (which can be a possible sign of an unfortunate accident) and notify persons preset by the user in the application, in the face of such an event. Context awareness can be exploited to trigger alarms and events based on sensor data, and can be extended to other applications such as virtually *"looking-after"* aged parents or handicapped individuals.

## V. System and Network Architecture

Technological advances in sensing, computation, storage, and communications are turning the near-ubiquitous mobile phone into a global mobile sensing device. The information exchange framework for mobile sensor networks using Android Phones require the ability to activate the sensors in the phones, send the sensor data to another phone and store this data for analysis and use in application areas such as personal, public and social sensing.

Therefore we propose the SenseDroid Architecture in support of such requirements. While incorporating the wireless Wi Fi infrastructure for the purpose of data transfer, the architecture primarily makes use of devices with embedded sensors like the mobile phones (specifically Android Phones). The main component of this architecture is the Central Server. The Central Sever is a physical computer dedicated for running a program that will receive, store and forward the queries and responses sent by the phones. The phones with the SenseDroid app installed register with the Central Server. The Server provides each phone a unique id(which can be a function of the phones number). So the Server has a table that maps each phone to its IP address and the unique id. Our architecture closely resembles the Client/ Server model. The architecture adopts an opportunistic paradigm where data collection is fully automated with no user interaction.

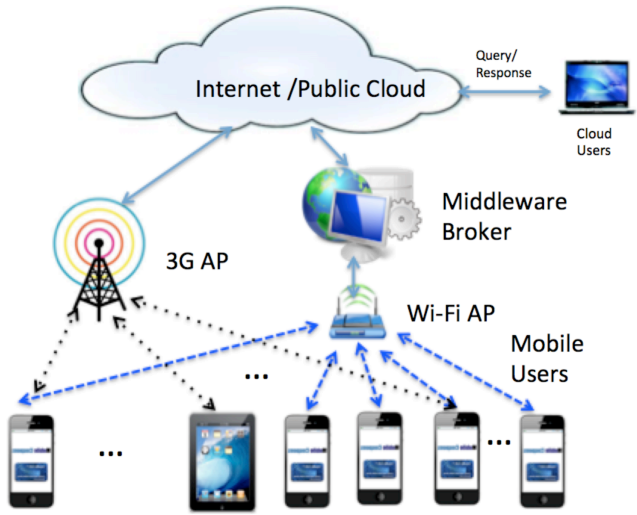

**Figure 3: SenseDroid System Architecture.**

To clearly understand the functioning of the various components in this architecture, let us consider a scenario where an Android Phone A wants to get the temperature of the place where another Android Phone B is located (as given in the figure above). The steps involved in this sensor information exchange are as follows:

a. Phone A (with IP address say IPA) registers itself with the server(assuming it knows the IP address of the server) using the WiFi. So the server now knows that phone A with IPA is online. Similarly, another phone B with IP address say IPB also registers itself with the server.

b. Now phone A wants to obtain the temperature of the place where phone B is located. So A sends a query to the server saying, 'Hey I want to know the temperature of the location where phone B is present '.

c. The server on receiving the query, checks if phone B is registered with it and thus is online.

d. If B is online, the server forwards the query to B. If B is not online, it sends a response to A saying that 'B is unavailable right now.'

e. Phone B, on receiving the query, processes the query by activating the appropriate sensor and reading the senor data. It prepares the reply saying, 'Hey X is the temperature of my location. ' It sends this reply to server asking it to forward the query response to A.

f. The server, on receiving the response, stores the response in its database and then forwards it to A.

Thus the communication between the two phones always goes through the Central Server. The power of this architectural choice is that it will allow information exchange between the phones connected to different WiFi networks even if the phones are located behind the NAT. The sensor information collected in the server can be used in applications described in the use cases above.

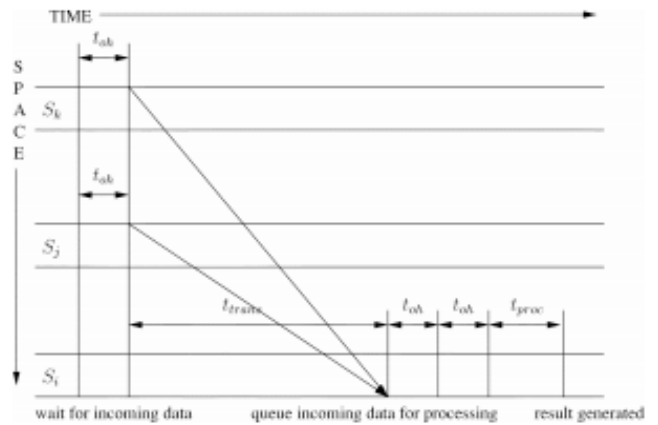

**Figure 4: Life cycle of Data in Proposed Architecture.**

## VI. Performance Analysis

The system and the application proposed could be qualitatively assessed using several performance parameters to maintain a quality-of-service (QoS). We consider the following two matrices to evaluate the performance of the client/server-based system:

### A. Execution Time

The execution time is the time spent to finish processing a task. In the proposed system, it starts from the time a query is generated to the time the node returns with results. For simplicity of discussion, we assume three sensors nodes $S_i, S_j$, and $S_k$ and assign $S_i$ as the processing center (i.e. $S_i$ as server $S_j, S_k$ as clients in the case of client/server-based computing ). In the client/server-based model, it is from the time the clients send out data to the time the data processing is finished and results are generated at the server. The execution time consists of three components, $t_{trans}$, $t_{oh}$, and $t_{proc}$, as illustrated in Fig. 3 where $t_{trans}$ is the time in transferring the sensor data from one node to the other, $t_{oh}$ represent the overhead time in accessing the data /file system, and $t_{proc}$ represents the processing time. A few factors that can affect the execution time include the network transfer rate $v_n$, the data processing $v_d$, the data file size $s_f$ (the size of raw data each node collects), the overhead of file access $o_f$ (the time used to read and write a data file), the number of sensor nodes $p$.

Thus, for client/server-based computing, the data transfer time is

$$t_{trans} = p \times s_f / v_n \qquad (1)$$

and the overhead time is

$$t_{oh} = 2p \times o_f \qquad (2)$$

(assuming the time used to read and write the data file is the same); and the data processing time is

$$t_{proc} = p \times s_f / v_d. \qquad (3)$$

Therefore, the total execution time using the client/server-based model is

$$t_{cs} = t_{trans} + t_{oh} + t_{proc}$$
$$= \frac{ps_f}{v_n} + 2po_f + \frac{ps_f}{v_d} \qquad (4)$$

In the analytical model described above, the component that is most difficult to measure is the data transfer time, where retransmission and error control are not considered. Unfortunately, these factors occur quite often in sensor networks because of the use of wireless link. Therefore, we develop simulation models for more accurate estimation of the data transfer time.

### B. Energy Consumption

Sensor nodes are normally composed of four basic units: a sensing unit, a processing unit, a communication unit, and a power unit. Among these units, communication and sensing consume most of the energy. However, since the energy consumed in sensing is the same for both models, we choose to neglect this factor.

Similar to the formulation of the execution time, the energy consumption for the two computing models depends on three components: energy consumed in data transfer $E_{trans}$, overhead processing $E_{oh}$, and data processing $E_{proc}$. Since no matter where the data processing takes place, be it at the local sensor node or the processing center, the energy consumed for the entire sensor network is the same for both computing models, we choose to neglect $E_{proc}$.

According to, the energy consumed in data transfer can be modeled using a linear equation,

$$E_{trans} = c \times size + d, \qquad (5)$$

where $d$ is a fixed component associated with device state changes and channel acquisition overhead, $size$ is the size of data being transferred, and $c$ is a coefficient indicating the amount of energy consumed by transferring 1 B of data. The values for $c$ and $d$ are different between data transmission and data receiving. Therefore, we separate into two parts,

$$E_{tx} = c_{tx} \times size + d_{tx}, \qquad (6)$$

for transmission and for receiving:

$$E_{tr} = c_{tr} \times size + d_{tr}. \qquad (7)$$

We use a similar linear equation to model the energy consumption in overhead processing,

$$E_{oh} = c_{proc} \times size \qquad (8)$$

where is the coefficient indicating the amount of energy consumed in processing 1 B of data. Since we only have knowledge of time spent for overhead processing and the amount of data that can be processed in 1 s, the so-called equivalent data size $s_e$, we can derive the size of data that takes $o_f$ amount of overhead time to process, that is, $s_e o_f$ for the client/server-based model. Since $s_e$ differs when the number of clients changes we choose an average value.

Similar to (4), the energy consumption model we use for client/server-based computing is given bellow.

$$E_{cs} = p[(c_{tx}s_f + d_{tx}) + (c_{rx} + d_{rx}) + 2(c_{proc}s_e o_f)] \qquad (9)$$

### VII. SUMMARY

This paper dealt with the state of the art in research and development of Mobile Phone Sensing Systems. The incremental addition of sensors into the mobile devices not only makes them smarter but holds the potential to revolutionize the conventional ways of collection and analysis of data, which impact various fields of human life directly and indirectly.

The paper starts off discussing the root causes behind the sudden upward surge in mobile sensing based applications followed by a brief introduction to the sensors present in most smartphones today. We also discuss how mobility in/during sensing greatly enhances the span of sensed area; paradigms of sensing - opportunistic sensing being the mode where minimal (to nil) user involvement is anticipated versus the user participated and initiated sensing. Various application areas where mobile sensing can (and does) play pivotal roles have been explained in adequate detail in the applications and use

S.Sarma, University of California, Irvine.

cases section. The existing architectures, their shortcomings and challenges; and the proposed network and software architecture of SenseDroid is dealt with, in great details. A performance analysis is presented that quantitatively (and in some sense qualitatively) treats the subject under consideration.

## REFERENCES

[1] I. Vasilescu, K. Kotay, D. Rus, M. Dunbabin, and P. Corke. Data collection, storage, and retrieval with an underwater sensor network.

[2] W. Kaiser, G. Pottie, M. Srivastava, G. Sukhatme, J. Villasenor, and D. Estrin. Networked Infomechanical Systems (NIMS) for Ambient Intelligence. Ambient Intelligence, 2004.

[3] Lane, N.D.; Miluzzo, E.; Hong Lu; Peebles, D.; Choudhury, T.; Campbell, A.T. "A survey of mobile phone sensing," Communications Magazine, IEEE , Volume: 48 , Issue: 9 , pp. 140 – 150, 2010.

[4] CarTel: A Distributed Mobile Sensor Computing System Bret Hull, Vladimir Bychkovsky, Yang Zhang, Kevin Chen, Michel Goraczko, Allen Miu, Eugene Shih, Hari Balakrishnan and Samuel Madden MIT Computer Science and Artificial Intelligence Laboratory

[5] Pasztor, B.; Musolesi, M.; Mascolo, C. "Opportunistic Mobile Sensor Data Collection with SCAR," Mobile Adhoc and Sensor Systems, 2007. MASS 2007. IEEE Internatonal Conference on , pp. 1 – 12, 2007.Hans-W. Gellersen1, Albrecht Schmidt and Michael Beigl "Multi-Sensor Context-Awareness in Mobile Devices and Smart Artefacts," Department of Computing, Lancaster University

[6] A. Kapadia, D. Kotz, and N. Triandopoulos, "Opportunistic Sensing: Security Challenges for the New Paradigm," Proc. 1st COMNETS, Bangalore, 2009.

[7] J. Healey and R. Picard, StartleCam: A Cybernetic Wearable Camera, Proc. of the International Symposium on Wearable Computing, Pittsburgh, PA, USA, October 1998

[8] A. Schmidt, Implicit Human-Computer Interaction through Context, Personal Technologies 4(2&3), June 2000

[9] M. Weiser, The Computer of the 21st Century. Scientific American 265, 3, September 1991.

[10] V. Tuulos, J. Scheible, and H. Nyholm, "Combining Web, Mobile Phones and Public Displays in Large-Scale: Manhattan Story Mashup," Proc. 5th Int'l Conf. Pervasive Computing, LNCS 4480, Springer, 2007

[11] http://research.nokia.com/page/232

[12] A. Kansal et al., "SenseWeb: An Infrastructure for Shared Sensing," IEEE MultiMedia, vol. 14, no. 4, 2007, pp. 8–13.

[13] A. Parker et al., "Network System Challenges in Selective Sharing and Verification for Personal, Social, and Urban-Scale Sensing Applications," Proc. 5th Workshop Hot Topics in Networks (HotNets-V), 2006.

[14] Andrew T. Campbell, Shane B. Eisenman, Nicholas D. Lane, Emiliano Miluzzo, Ronald A. Peterson, Hong Lu, Xiao Zheng, Mirco Musolesi, Kristóf Fodor , and Gahng-Seop Ahn, "The Rise of People-Centric Sensing", In IEEE Internet Computing: Mesh Networking, pp. 30-39, July/August, 2008.

[15] M. Ho and K. Fall. Delay Tolerant Networking for Sensor Networks (Poster Abstract). In Proc. IEEE 1st Conf. on Sensor and Ad Hoc Communications and Networks (SECON '04).

[16] Y. Wang and H. Wu. DFT-MSN: The Delay Fault Tolerant Mobile Sensor Network for Pervasive Information Gathering. In Proc. IEEE 25th Int'l Conf. on Computer Communications (INFOCOM'06), 2006.

[17] A. Arora, et al. Exscal: Elements of an extreme scale wireless sensor network. In Proc. IEEE 11th Int'l Conf. on Embedded and Real-Time Computing Systems and Applications (RTCSA), pp 102–108,2005.

[18] 0. Gnawali, K. Jang, J. Paek, M. Vieira, R. Govindan, B. Greenstein, A. Joki, D. Estrin and E. Kohler. The Tenet Architecture for Tiered Sensor Networks. In Proc. ACM 4th Int'l Conf. on Embedded Networked Sensor Systems (SENSYS '06), pp 153–166, Boulder, Oct/Nov 2006.

[19] C.-Y.Wan, S. B. Eisenman, A. T. Campbell and J. Crowcroft. Overload TrafficManagement for Sensor Networks. In ACM Trans. on Sensor Networks, Vol 3, No 4, pp 1–38, 2007.

[20] F. Stann and J. Heidemann. RMST: Reliable Data Transport in Sensor Networks. In Proc. 1st IEEE Int'l Workshop on Sensor Net Protocols and Applications, pp 102–112, Anchorage, Apr 2003.

[21] Khan, W.; Xiang, Y.; Aalsalem, M.; Arshad, Q. "Mobile Phone Sensing Systems: A Survey," Communications Surveys & Tutorials, IEEE Volume: PP, Issue: 99 , pp. 1 – 26, 2012.

[22] Hairong Qi; Yingyue Xu; Xiaoling Wang "Mobile-agent-based collaborative signal andinformation processing in sensor networks ," Proceedings of the IEEE , Volume: 91 , Issue: 8 , pp. 1172 – 1183, 2003.